# Reasons behind growing adoption of Cloud after Covid-19 Pandemic and Challenges ahead

Mayank Gokarna

IBM India Pvt Ltd, Manyata Tech Park, Bangalore, India
Email: mayank123manit@gmail.com

## Abstract

There are many sectors which have moved to Cloud and are planning aggressively to move their workloads to Cloud since the world entered Covid-19 pandemic. There are various reasons why Cloud is an essential irresistible technology and serves as an ultimate solution to access IT software and systems. It has become a new essential catalyst for Enterprise Organisations which are looking for Digital Transformation. Remote working is a common phenomenon now across all the IT companies making the services available all the time. Covid-19 has made cloud adoption an immediate priority for Organisation rather than a slowly approached future transformation. The benefits of Cloud lies in the fact that employees rather engineers of an enterprise are no more dependent on the closed hardware-based IT infrastructure and hence eliminates the necessity of working from the networked office premises. This has raised a huge demand for skilled Cloud specialist who can manage and support the systems running on cloud across different regions of the world. In this research, the reasons for growing Cloud adoption after pandemic Covid-19 has been described and the challenges which Organization will face is also explained. This study also details the most used cloud services during the pandemic considering Amazon Web Services as the cloud provider.

## Introduction

Cloud Computing is no more a new word in IT industry as it has been there since many years now but what is Cloud computing and why its adoption among IT organizations has been increasing at a much faster rate than ever? Why pandemic like Covid-19 has forced certain Organizations to think and plan to adopt Cloud computing.? Cloud computing is the availability of various computing services which includes primarily compute, storage and network over the internet to enable flexibility to access, faster innovation, lower infrastructure costs and making your end user systems highly available and scalable with increased security. It reduces the burden of maintaining the on-premise infrastructure and since it's an on-demand delivery model you pay-as-you-go.

## Essential benefits of Cloud during Pandemic situation

### 1, Remote Working Solution

Getting the work done from anywhere is remote working. This practise has been there since years with some of the large enterprise companies but the real benefit or the importance of it was realised when the workforce across the world was forced to stay home post Covid-19 outbreak in March 2020 as the lockdown imposed was the only way appeared to stop the spread of the virus. The pandemic affected every aspect of life at every level and corporate work culture was not an exception however enterprises with Cloud based infrastructure and capabilities were able to sustain this drastic change at workplace since they had the remote working solution in practise. Not only some but a huge number of enterprises started leveraging the benefits of cloud adoption during this pandemic situation and the rest of the businesses started realising that Cloud capabilities are a solution to their problems [1].

## 2, Business continuity

Enterprises with robust IT infrastructure based on Cloud were able to function well even in the times of pandemic when each and every business was getting adversely affected. Secured and quick access to protected data is an important factor to continue business. Having data stored in the cloud always ensures that it is safe and secured irrespective of the location from where you are accessing it [2].

With pandemic spread across the world since March 2020, various Governments started enforcing lockdown measures to the entire cities, towns and countries. With this different business started getting adversely impacted as the physical appearance of employees was restricted with only a very few exceptions. This triggered the surge in the use of virtual tools and techniques to stay connected with the business within this short time frame. Various online meeting platforms like Google Meet, Zoom, Microsoft Teams, Webex, etc reported sudden increase of revenue specifically with Zoom being more than 100% increase in revenue due to the sudden surge in the use of their digital virtual technologies [3].

## 3, Efficient Collaboration

Any business having more than few employees need to make an essential daily collaboration in order to perform as an efficient team. Its rather one of the top priorities to keep the team members collaborated and connected during the working hours. Cloud made this collaboration much simpler in terms of the complexity it delivers. They can connect over video conferences, share content and information, discuss in a group securely over a cloud-based platform. Collaboration may be possible without a cloud computing solution but keeping this collaboration easy and seamless is an important factor when different employees work from different geographical locations.

If your business has two employees or more, then you should be making collaboration a top priority. After all, there isn't much point to having a team if it is unable to work like a team. Cloud computing makes collaboration a simple process. Team members can view and share information easily and securely across a cloud-based platform. Some cloud-based services even provide collaborative social spaces to connect employees across your organisation, therefore increasing interest and engagement. Collaboration may be possible without a cloud-computing solution, but it will never be as easy, nor as effective as with a Cloud based solution. This is done in order to improve efficiency and performance.

## 4, Remote Education

Covid-19 pandemic has forced Government authorities to lock down schools and colleges and the students are made to study at home by virtual means. Educators have enabled virtual solutions for teachers and students in the wake of pandemic by implementing virtual solutions [23] like Microsoft Teams which not only provides video-enabled remote classes but also a platform to study and share documents and homework according to the age groups[4]. With these platforms teachers are able to develop closer connection with their students. It is a convenient way of attending the online classes from any place and if needed recordings can help to review the content taught in the live online class. With video enabled e-learning platforms teachers and students are kept engaged during the classes and parents too can see the attention and performance shown while attending the online classes [5].

## 5, Cybersecurity

With Cloud computing the infrastructure belongs to the Cloud provider and is managed and administered by them. The Cloud adopting companies or Organizations always have certain level of concerns over the security of their data and files since they do not know where and how secured is their data with the cloud provider. Public and private organizations have invested millions in security products and incident response personnel but have not tested the efficiency of these safeguards and people using actual attack techniques. They generally lack a comprehensive approach to understanding and operationalizing cyber threat intelligence. The result is a weak security defence posture that undermines investments in threat detection and protection. But Cloud providers have ascertained that they monitor the Client's data and files all the times with latest security policies and services. With a shared responsibility model followed on cloud, it is essential for an enterprise to monitor, identify and remediate on any potential threats and misconfigurations on their provisioned cloud resources. Data breaches and improper Identity and Access management are the important factors to be considered before going for cloud adoption [6]. There are certain obvious common security concerns as well with both on-premise and Cloud based infrastructure. With Cloud adoption, RapidScale claims that 94% of businesses saw an improvement in security and 91% said the cloud makes it easier to meet government compliance requirements [7].

## 6, Scaling

Cloud based Infrastructure is a great solution for enterprises as it enables them to scale up/down their IT resources efficiently and quickly, according to the business demands [19]. It is ideal for enterprises with fluctuating demands. As demands rise the cloud will provision the needed infrastructure and services without having to bother about the physical infrastructure. This agility of cloud enables an enterprise to manage their cost efficiently and without keeping any of the resources idle. Organisation that adopt Cloud solutions always has a real advantage over their competitors.

Cloud based solutions are ideal for businesses with growing or fluctuating bandwidth demands. If your business demands increase, you can easily increase your cloud capacity without having to invest in physical infrastructure. This level of agility can give businesses using cloud computing a real advantage over competitors [8].

The auto scaling capabilities of Cloud eliminates the risks associated with on-premise infrastructure operational and maintenance issues. One of the biggest advantages of it is that it has no upfront cost involved and you can control your Cloud expenditures as per your business and budget.

## 7, Everlasting advantages of Cloud adoption

While Cloud was born to provide certain essential features like High Availability, Scalability, Business Continuity, Fault Tolerance, Disaster Recovery, Automatic Software Updates, Flexibility etc, over the time, a lot of new features and services have emerged with the growing interest in the adoption of Cloud by various enterprises. Originally Amazon Web Services emerged as a sustainable Cloud provider but now there are various other Cloud providers too like Microsoft's Azure, Google Cloud, IBM Cloud, Oracle etc. Each of these providers have an edge over the other Cloud provider in terms of different areas like Network, Storage, Security, Compute, Availability etc.

## Most used Cloud Services during Covid-19

Enterprises using Cloud for hosting their applications to be accessible over internet securely used a variety of Cloud services. The table below in Figure 1. shows Cloud services which were used most by 4 different Organisation using AWS (Amazon Web Services) as the Cloud Provider for their 3 applications.

| S.No. | AWS Cloud Services | Service type |
|---|---|---|
| 1. | S3 | Simple Storage Service |
| 2. | EC2 | Virtual Machines |
| 3. | Cloudfront | Global Content Delivery Service |
| 4. | RDS | Relational Database Service |
| 5. | SNS | Simple Notification Service |
| 6. | IAM | Identity and Access Management |
| 7. | VPC | Virtual Private Cloud |
| 8. | AutoScaling | Scaling Service |
| 9. | Elastic Load Balancers | Traffic Distribution Service |
| 10. | Beanstalk | Platform as a Service |
| 11. | Lambda | Serverless Computing Platform |

Figure 1. Most used AWS Cloud services

## Types of Clouds

**Public Cloud:** It's a type of hosting in which cloud services are delivered over a network for public use, in a shared Infrastructure model. Customers do not have any control over the location of the infrastructure in Public Cloud and the cost is shared and are either free or in the form of a license policy like pay per user. Public Clouds are great for organizations that require managing the host applications

**Private Cloud:** It's a Cloud infrastructure that is solely used dedicatedly by one organization. It gives Organizations greater control over security and data which is managed internally. It can be hosted internally or externally. Private clouds are great for organizations that have high security demands, high management demands and uptime requirements.

**Hybrid Cloud:** It uses both private and public clouds but can remain separate entities or isolated from each other [26]. Resources are managed and can be provided either internally or by external providers. A hybrid cloud is great for scalability, flexibility and security.

**Community Cloud:** A Cloud computing model with a collaborative effort in which infrastructure is shared between different organizations of a specific community with common objectives and managed internally or by a third-party and the Cloud can also be hosted internally or externally.

## Cloud computing operational Models:

Figure 2. below shows different Cloud Service Models with the level of administration possible at each of the Models and the type of user responsible for managing the administration.

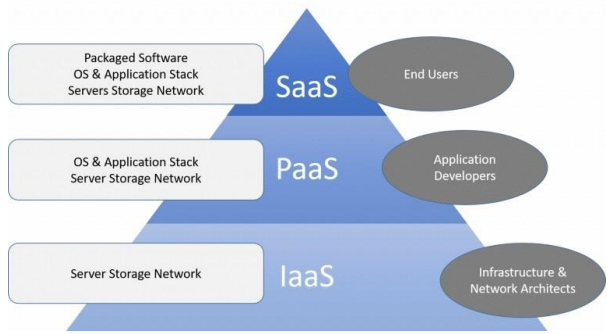

Figure 2. Cloud Computing Operational Service Models

### SaaS: Software as a Service

SaaS services are accessible over internet on a browser and is a quick way to get your business leverage these services for use according to requirements [21]. SaaS services are managed by the vendors and frees up the technical staff of your business which could otherwise be needed to manage it. It doesn't require any application or software to be downloaded at the client side and is easily accessible [20]. Examples of SaaS services are dropbox, salesforce, GoToMeeting [9].

### PaaS: Platform as a Service

PaaS services is a delivery of a framework which a business can use to further build upon it the application based on the customizations needed [25]. It gives a base platform quickly to start building your application upon. While developers are responsible for their own custom application, the servers, storage and networking of the platform is managed by the third-party service provider. Examples of PaaS services are Google App engine, AWS Elastic Beanstalk, Openshift etc.

### IaaS: Infrastructure as a Service

IaaS services is a delivery of cloud computing infrastructure which includes network, storage, servers, operating systems etc. It is one of the most flexible cloud computing models in which resources are available as services. In this model, the complete control over the infrastructure lies with the user Organization. It is highly scalable and dynamic. Examples of IaaS services are Amazon Web Services EC2 instances, Google Cloud Compute Engine, Cisco Metacloud etc.

## Cloud is the new normal

Cloud has been evolving since over 2 decades and the enterprise arguments on its security and resiliency has been proven time and again as one of the largest U.S. Defence Department project called JEDI worth $10 billion has been bagged by Microsoft's Azure [10]. Seeing the Cloud adoption benefits, it was evident that Cloud will become the future enterprise technology infrastructure and all organisations need to slowly embrace the cloud [17].

However, there has been compressed cloud adoption curve since the pandemic Covid-19 started back in March 2020. A lot of Organisations already experienced and realised the power and benefits of Cloud to make their business operations more resilient and highly available.

Cloud computing has been playing an increasing role in ensuring the seamless provision of services, and this has been demonstrated throughout the Covid-19 pandemic, not only this but it has also enabled the opportunity to provide additional new services in seamless fashion [11]. In response to Covid-19, in the month of April 2020, Google made their online meeting platform Google Meet to be free for a duration of first 60 minutes in the free version of this SaaS application. Usage of lot of other videoconferencing tools has also increased drastically since the Covid-19 outbreak.

Microsoft Teams and Zoom have also seen the daily number of users skyrocket since then and Google Meet is no exception. Everyday use of such SaaS application video conferencing tools is now thirty times higher than before Covid-19. Millions of users are increasingly joining these services to keep their businesses up and running [12].

Data prepared by US Department of Labour, O'Net Center shows which occupations carry the most risk of infection from Covid-19. Below Figure 3. shows the visual representation of the study conducted [13]. This study shows that IT occupations such as computer engineers, architects, programmers, developers fall in the lowest risk percentage ranging from 0-20%. This obviously is attributed to the remote working capabilities provided in the IT infrastructure adopted by most Organizations. A further increase in digital transformation and Cloud adoption can further bring down the risk factor caused by Covid-19 infections. This is a signal to prepare better for future pandemic situations.

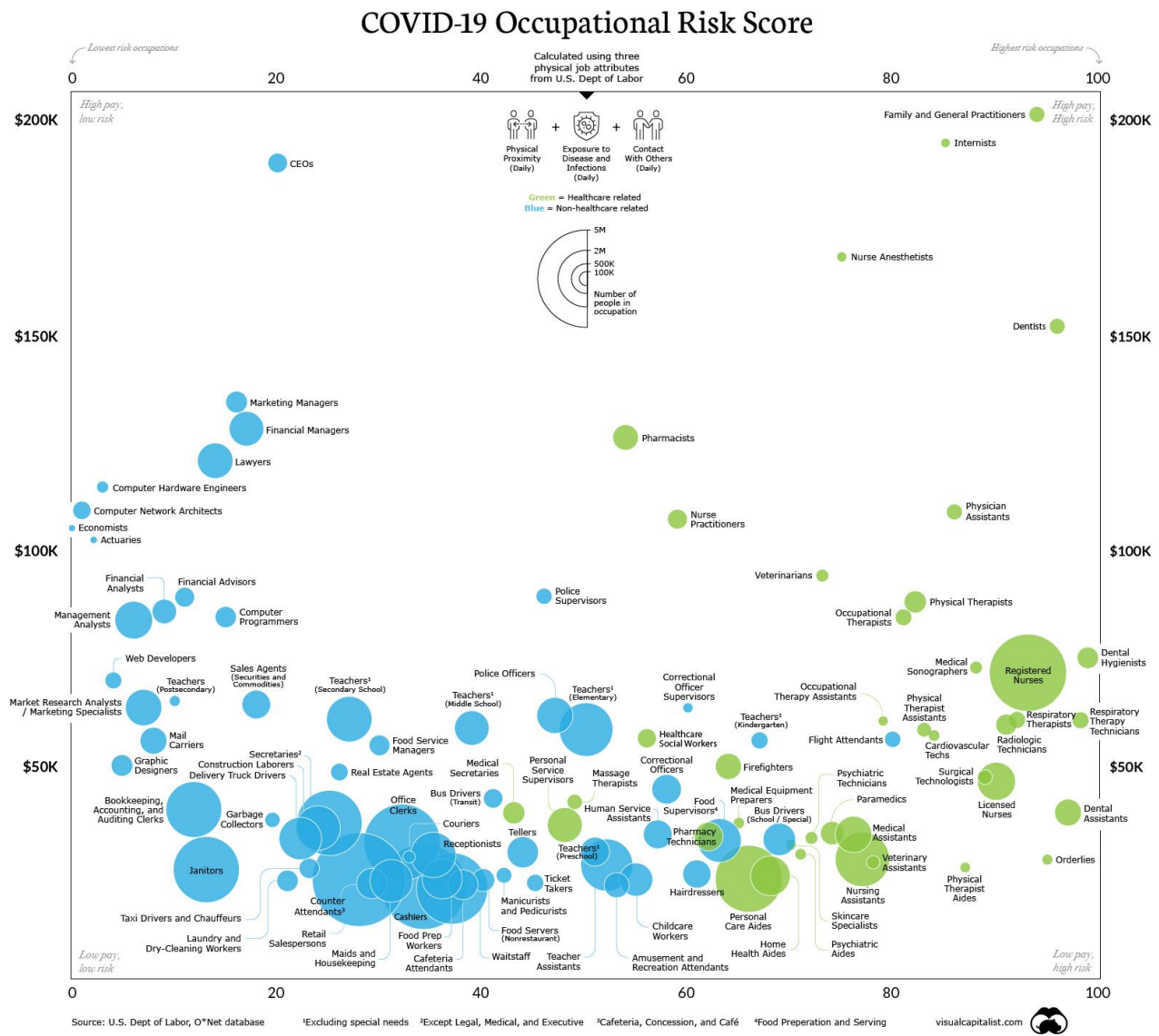

Figure 3. Various Occupational Risk score

## Challenges ahead

Though there are numerous advantages of Cloud adoption as compared to on-premise infrastructure, however there are some obvious challenges which should be taken care over the period of time in this emerging trend. Australian Government has published various risk issues posed by the adoption of Cloud and a need to conduct a risk assessment before going to decide on the shared responsibilities of Cloud Provider and the end user [14].

### 1, Security and Data Protection

It is critical to decide on how the data will be stored and what security features will be applied to it before giving control to the Cloud provider [22]. The extent of security depends on the level of confidential data you currently own. The measures to ensure data security and how it will be accessed should be clearly defined and made known to the Organization by the service provider [18]. Data at transit and at rest should be encrypted and the level of encryption can be decided mutually [27]. Data security in the cloud world is more complicated than data security in the on-premise or traditional information systems for the fact that data is scattered at different places in different devices.

### 2, Data Location

It is important to know where the data will physically reside as there are privacy and security laws that apply to the data in many countries. Sometimes the data is required to be made present at more than one continent to make it highly available [15] however the local privacy and security laws cannot be ignored.

### 3, Skills shortage

Hiring and retaining talent with strong Cloud skills is another burden for Cloud adoption. Any cloud migration project introduces new capacities and benefits over the traditional architecture however if it's not done properly then it may turn into a source of unexpected vulnerabilities. Having skilled and Cloud certified professionals in the Organization to kick off the cloud journey process is of utmost importance and scarcity of such engineers is a huge challenge to continue with this journey.

## Conclusion

It is evident during Covid-19 that Cloud is an essential enabler for remote data access, storage and perform operations from remote location which ensured business continuity and risk mitigation for the production systems. It is also seen that Organisations with strong Cloud capabilities performed much better compared to the Organisation with on-premise infrastructure during these pandemic times.

Many Organisations have accelerated their Cloud journey to a much faster pace than ever. The Covid-19 experience has highlighted the Cloud's value as a crucial element for risk mitigation for the businesses. It has not only provided service continuity but has also enabled remote working capabilities for the Organizations and eliminated the need to be physically present in an office workplace.

Many large and mid-size enterprises are reviewing their future mode of work strategy as Cloud has been considered as a key enabler to define it. Cloud comes with certain new responsibilities of obeying privacy and security laws [24] of the local region and ensuring proper security mechanism including encryption and vulnerability management. With some known challenges it is inevitable to adopt Cloud for new projects and migration of legacy applications [16] as the current pandemic situation has made the future more uncertain and volatile for business continuity.

## Availability of data and material

## Competing interests

## Funding

## Authors' contributions

Mayank Gokarna has written this paper and have done the research which supports it.

## Authors' information and details

Mayank Gokarna is a Lead DevOps and Cloud Architect at IBM India Pvt. Ltd. He is a Google certified Professional Cloud Architect and Engineer, AWS Certified Solutions Architect with a decade of experience working on DevOps tools and technologies. He has immense interest in implementing DevOps and best practices around it defining the Governance and maturity models. He has onboarded more than 50 applications on cloud platform using various Cloud managed and unmanaged services. He specializes in IaaS, PaaS and SaaS services. He received his BTech in Electronics & Communication engineering from National Institute of Technology, Bhopal.

## Acknowledgements

I thank Cloud Consultant Neetu Singh from IBM, India for her support in gathering data from different Organizations for mostly used cloud services using AWS (Amazon Web Services) Cloud provider.